\documentclass[10pt,english]{article}
\usepackage{mathptmx}
\usepackage[T1]{fontenc}
\usepackage[latin9]{luainputenc}
\usepackage{babel}
\usepackage{amsmath}
\usepackage{amssymb}
\usepackage{graphicx}
\usepackage{setspace}
\usepackage[pdfusetitle,
 bookmarks=true,bookmarksnumbered=false,bookmarksopen=false,
 breaklinks=false,pdfborder={0 0 1},backref=false,colorlinks=false]
 {hyperref}

\makeatletter

\newcommand{\lyxmathsym}[1]{\ifmmode\begingroup\def\b@ld{bold}
  \text{\ifx\math@version\b@ld\bfseries\fi#1}\endgroup\else#1\fi}

\newcommand{\lyxaddress}[1]{
	\par {\raggedright #1
	\vspace{1.4em}
	\noindent\par}
}

\makeatother

\begin{document}
\title{P--adic numbers and kernels}
\author{Simone Franchini}
\date{~}
\maketitle

\lyxaddress{\begin{center}
\textit{CNR-ISTC, Via Gian Domenico Romagnosi 18, 00196 Rome, Italy}
\par\end{center}}

~
\begin{abstract}
We discuss the relation between p--adic numbers and kernels in view
of a recent large deviation theory for mean-field spin glasses. As
an application we show several fundamental properties of numerical
bases in kernel language. In particular, we show that Derrida's
Generalized Random Energy Model can be interpreted as a (random) numerical
base. We also show an application to the Primon gas and the Riemann
Zeta Function by constructing a kernel representation of the Primon
gas based on a finite $p-$base, thereby establishing a concrete link
between number theory and kernel theory.

~


~

~
\end{abstract}
In what follows we review the discoveries of Parisi, Sourlas \cite{ParisiSourl}
and Avetisov, et al. \cite{ParisiSourl2} on the relations between
p--adic numbers \cite{Krennikov,Aniello,Zuniga-Galindo} and the
Replica Symmetry Breaking (RSB) theory \cite{PMezVir} in the light
of a recently introduced kernel method \cite{FranchiniRSBwR2023,Franchini2017,SimplifiedParisiIII,Franchini2024,BardellaFranchini2024,BardellaFranchiniShort2024}. 

Our main result will be to apply the known links between kernels and
RSB \cite{FranchiniRSBwR2023,Franchini2017,SimplifiedParisiIII} and between
RSB and number theory \cite{Varadarajan,Varadarajan2,Marek=000020Wolf}
to deduce a kernel representation for the p--adic numbers, and link
number theory with the kernel theory described in \cite{FranchiniRSBwR2023,Franchini2017}. 

In particular, we show that the Generalized Random Energy Model, (GREM)
\cite{DerridaGREM,Kistler}, although being random, constitutes a
legitimate bijective map for any numerable set of consecutive integers.
Finally, we explore the connection with the Riemann Zeta Function
(RZF) \cite{Schumayer,Berry,Fyodorov,Arguin,Sierra,Bender-Brody-Muller,Moxley}
by applying our findings to a classic result by Spector and Julia
\cite{Julia,Spector}

\section{Binary numbers and kernels}

\begin{figure}
\centering{}\includegraphics[scale=0.3]{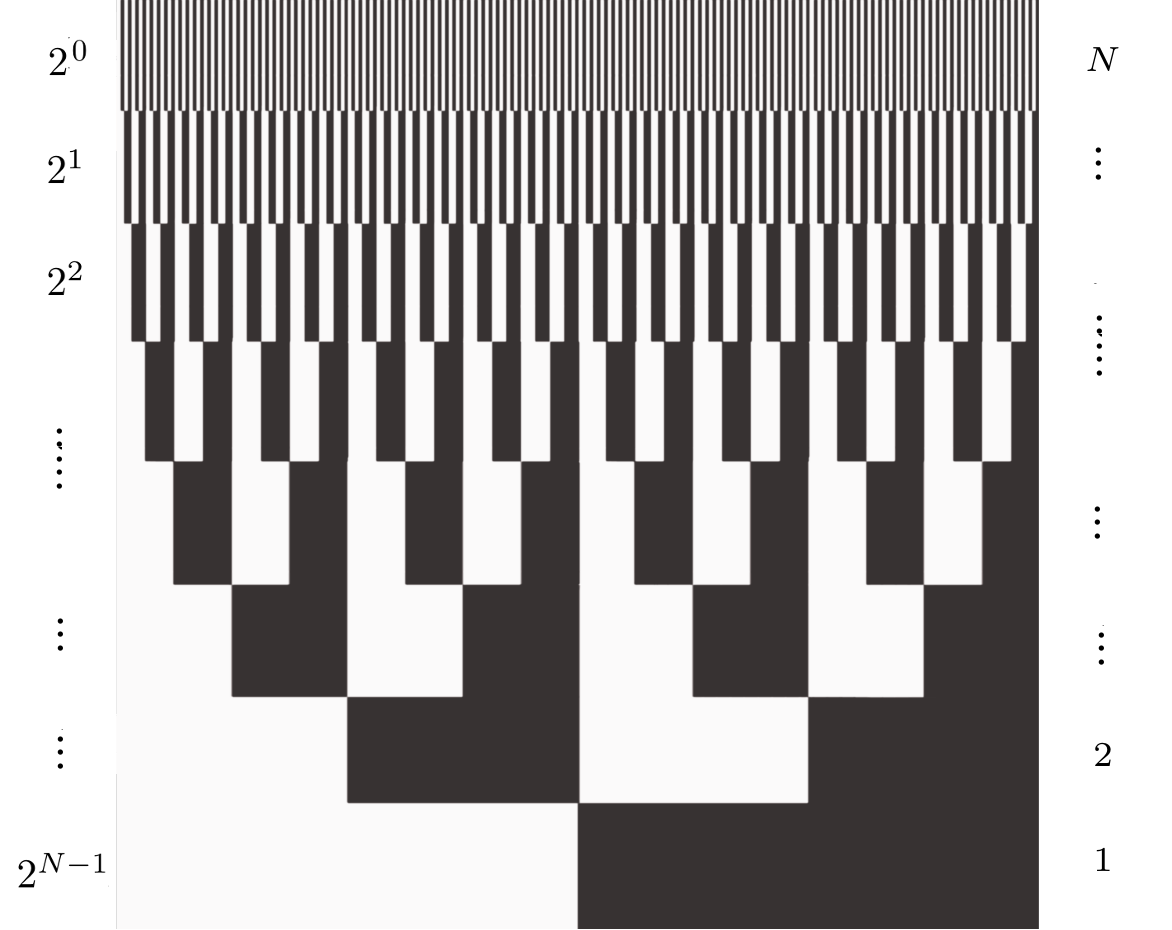}\caption{\protect\label{fig:Binary-kernel-describing}Binary kernel describing
the first $2^{8}=256$ natural numbers $\mathbb{N}_{256}$ (zero included).
The numbers are organized as columns of the kernel, ordered from the
smallest number, $0$, on the left to the largest, $2^{8}-1=255$, on the far
right. The index $i$ runs from $1$ at the bottom to $N$ at the top (shown on
the right). The corresponding base vector to construct the map $M_{2}$
is shown on the left. }
\end{figure}

\noindent Let $n\geq1$ be a natural number, and let 
\begin{equation}
\mathbb{N}_{n}=\left\{ 0,1,2,\,...\,,n-1\right\} \subset\mathbb{N}
\end{equation}
be the ordered set of the first $n$ nonnegative integers. Then let us introduce the notation $\underline{a}$ for the generic
$N$-digit number in the binary base 
\begin{equation}
\underline{a}=a_{1}\,a_{2}\,...\,a_{N}\in\mathbb{N}_{2}^{N},
\end{equation}
where $a_{i}\in\mathbb{N}_{2}=\left\{ 0,1\right\} $ is the $i-$th
binary digit. For example, the number $10$ in decimal notation is
written as $1010$; we say that $\left(10\right)_{10}=\left(1010\right)_{2}$.
There is a map between the above patterns and the natural numbers
$m$ from $0$ up to $2^{N}-1$
\begin{equation}
\mathbb{N}_{2}^{N}\ni\underline{a}\ {\textstyle \overset{M_{2}}{\longrightarrow}}\ m\in\mathbb{N}_{2^{N}}
\end{equation}
\begin{equation}
\mathbb{N}_{2}^{N}\ni\underline{a}\ {\textstyle \overset{M_{2}^{-1}}{\longleftarrow}}\ m\in\mathbb{N}_{2^{N}}
\end{equation}
It is easy to obtain a correspondence of $\underline{a}$ with the
spin $\Omega=\left\{ -1,1\right\} $ systems by taking $\sigma_{i}=2a_{i}-1$
for any $\sigma\in\Omega^{N}$. The map $m\left(\underline{a}\right)=M_{2}\left(\underline{a}\right)$
from $\underline{a}$ to $m$ is simply given by the following formula:
\begin{equation}
M_{2}\left(\underline{a}\right)=2^{N-1}a_{1}+2^{N-2}a_{2}+\,...\,+2^{N-i}a_{i}+...\,+a_{N}
\end{equation}
from which we could write the first $2^{N}$ natural numbers. We can
arrange the possible choices of the vector $\underline{a}$ as the
columns of some array, with $N$ rows and $2^{N}$ columns: hereafter,
we will call \textquotedbl kernel\textquotedbl{} such an array, see
Figure \ref{fig:Binary-kernel-describing}. This object can be related
to probability measures, operators and graphs (it is a graphon) \cite{FranchiniRSBwR2023,Franchini2017,SimplifiedParisiIII,Franchini2024,BardellaFranchini2024,BardellaFranchiniShort2024}.
Also, we could define a kernel rescaled by $2^{-N}$,
which is shown in Figure \ref{fig:Rescaled-kernel,-the}. In Figure
\ref{fig:We-can-also} we show a schematic reversal of the hierarchical
weights, inspired by the formal continuation discussed in \cite{ParisiSourl}. 

\begin{figure}
\centering{}\includegraphics[scale=0.14]{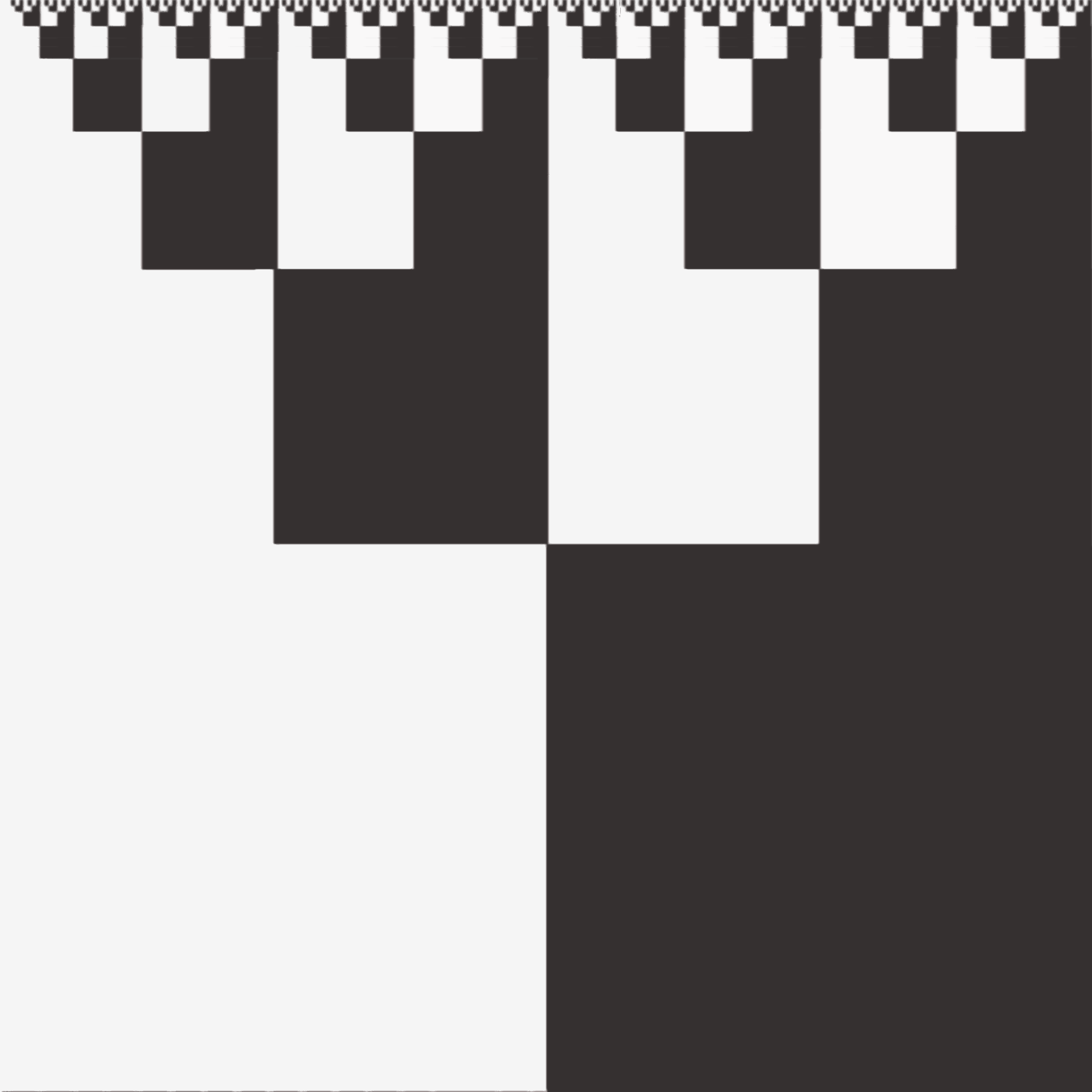}\caption{\protect\label{fig:Rescaled-kernel,-the}Rescaled kernel: the weighted sum
of each column is exactly the original number rescaled by the
factor $2^{-8}=1/256$. The numbers are ordered from $0$ on
the left to $255$ on the right. }
\end{figure}

\begin{figure}
\centering{}\includegraphics[scale=0.14]{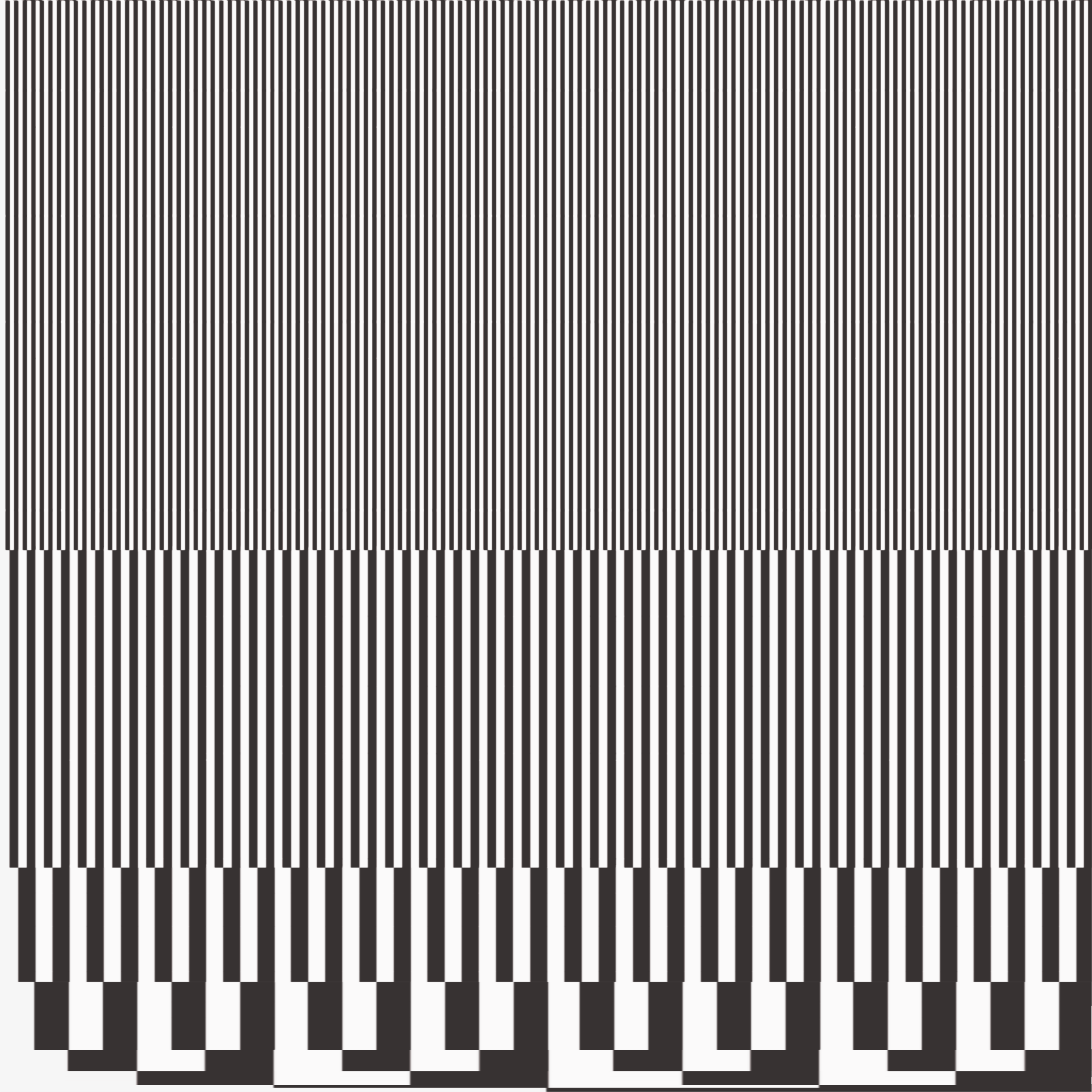}\caption{\protect\label{fig:We-can-also}Dual kernel obtained by reversing the geometric weights of the rows,
providing a schematic illustration inspired by the formal continuation
discussed in \cite{ParisiSourl}.}
\end{figure}

\section{Extension to $p-$adic numbers}

Now we generalize to $p-$adic numbers, where we assume $p$ to be
a prime number. The number is represented by the pattern
\begin{equation}
\underline{a}=a_{1}\,a_{2}\,...\,a_{N}\in\mathbb{N}_{p}^{N},
\end{equation}
where $a_{i}\in\mathbb{N}_{p}=\left\{ 0,1,2\,...\,,p-1\right\} $
is the $i-$th $p-$adic digit. Also in this case there is a bijective
map between the above patterns and the natural numbers from $0$ to
$p^{N}-1$
\begin{equation}
\mathbb{N}_{p}^{N}\ni\underline{a}\ {\textstyle \overset{M_{p}}{\longrightarrow}}\ m\in\mathbb{N}_{p^{N}}
\end{equation}
\begin{equation}
\mathbb{N}_{p}^{N}\ni\underline{a}\ {\textstyle \overset{M_{p}^{-1}}{\longleftarrow}}\ m\in\mathbb{N}_{p^{N}}
\end{equation}
The map $m\left(\underline{a}\right)=M_{p}\left(\underline{a}\right)$
from $\underline{a}$ to $m$ is 
\begin{equation}
M_{p}\left(\underline{a}\right)=p^{N-1}a_{1}+p^{N-2}a_{2}+\,...\,+p^{N-i}a_{i}+...\,+a_{N}
\end{equation}
from which we can write the first $p^{N}$ natural numbers. The tree
hierarchy is identified with the $p-$adic distance by reversing the
order of the digits relative to Eq. (9), as in \cite{ParisiSourl}.
For distinct patterns, this gives
\begin{equation}
\left|{\textstyle \sum_{i=1}^{N}}p^{i-1}\left(a_{i}-b_{i}\right)\right|_{p}=p^{-Q\left(\underline{a},\underline{b}\right)},\qquad\underline{a}\neq\underline{b}
\end{equation}
where $0\leq Q\left(\underline{a},\underline{b}\right)\leq N$ is
the number of consecutive congruent digits, starting from $i=1$ (ultrametric
index, see \cite{FranchiniRSBwR2023,Franchini2017,SimplifiedParisiIII}).
Let $\mathbb{I}\left(X\right)$ be the indicator function of the event
$X$; it equals one if the event occurs and zero otherwise. Then
\begin{equation}
Q\left(\underline{a},\underline{b}\right)={\textstyle \sum_{i=1}^{N}\prod_{j=1}^{i}}\,\mathbb{I}\left(a_{j}=b_{j}\right).
\end{equation}
It has been noted in \cite{ParisiSourl} that, since the $p-$adic
norm is ultrametric, it can be used to describe the RSB ansatz. Together
with the map $M_{p}$ defined above, this suggests that there is convenience
in interpreting numbers as patterns on a tree, whose branching ratio
is taken here as the fixed base $p$.

\section{Example: Random Energy Model}

We can actually build a representation for $\mathbb{N}_{n}$ where
the map is random: for simplicity we discuss this extension starting
from the binary base only (i.e., dyadic base $p_{i}=2$). Given $\underline{a}\in\mathbb{N}_{2}^{N}$
we introduce a generalized map 
\begin{equation}
M_{A}\left(\underline{a}\right)=A\left(\underline{a}_{1}\right)+A\left(\underline{a}_{2}\right)+\,...\,+A\left(\underline{a}_{N}\right)
\end{equation}
where we call $A\left(\underline{a}_{i}\right)$ the $i-$th generalized
binary digit. For example, we recover the binary system by choosing
$A\left(\underline{a}_{i}\right)=2^{N-i}a_{i}$ and $a_{i}\in\mathbb{N}_{2}=\left\{ 0,1\right\} $
(see Figures 4 and 5). In general, we expect that starting from any
set of $A\left(\underline{a}_{i}\right)\in\mathbb{R}$ such that 
\begin{equation}
M_{A}\left(\underline{a}\right)\neq M_{A}\left(\underline{b}\right)
\end{equation}
for any $\underline{a}\neq\underline{b}$ it is possible to produce
a bijective map to the first $2^{N}$ naturals (eventually this can
be extended to any kernel parametrization). Now, what if we chose
a random base system? For two distinct patterns, the common terms cancel and the difference
of their images is a nondegenerate Gaussian, so coincidences have probability zero. Then, 
\begin{equation}
A\left(\underline{a}_{i}\right)=J_{\underline{a}_{i}}=J_{a_{1}a_{2}...a_{i}}
\end{equation}
actually admits a valid reconstruction map 
\begin{equation}
\mathbb{N}_{2}^{N}\ni\underline{a}\ {\textstyle \overset{M_{A}}{\longrightarrow}}\ m\in\mathbb{R}
\end{equation}
\begin{equation}
\mathbb{N}_{2}^{N}\ni\underline{a}\ {\textstyle \overset{M_{A}^{-1}}{\longleftarrow}}\ m\in\mathbb{R}
\end{equation}
almost surely for each instance of the disorder. In the above expression
$J_{\underline{a}_{i}}$ are independent Gaussian random variables, one for each
$\underline{a}_{i}$, centered on $0$ and of variance $\gamma_{\underline{a}_{i}}^{2}$.
The noise has then the following covariance matrix
\begin{equation}
\mathbb{E}{\textstyle \,(J_{\underline{a}_{i}}J_{\underline{b}_{j}})}=\gamma_{\underline{a}_{i}}^{2}\mathbb{I}\left(i=j\right)\mathbb{I}{\textstyle \left(\underline{a}_{i}=\underline{b}_{i}\right)},
\end{equation}
where $\mathbb{I}\left(X\right)$ is the indicator function of the
event $X$ (that is one if $X$ is verified and zero otherwise). The
mapping is random and noise dependent 
\begin{equation}
M_{J}\left(\underline{a}\right)=J_{a_{1}}+J_{a_{1}a_{2}}+\,...\,+J_{a_{1}a_{2}...a_{i}}+\,...\,+J_{a_{1}a_{2}...a_{N}}
\end{equation}
and it is possible to recognize that this map is exactly the Hamiltonian
of the Generalized Random Energy Model (GREM) \cite{DerridaGREM,Kistler}.
The interesting feature of this base is that the averaged product
of two numbers is equal to the overlap; most importantly, the product
matrix is ultrametric. In fact
\begin{multline}
M_{J}\left(\underline{a}\right)M_{J}\left(\underline{b}\right)=\\
=\left(J_{\underline{a}_{1}}+\,...\,+J_{\underline{a}_{N}}\right)\left(J_{\underline{b}_{1}}+\,...\,+J_{\underline{b}_{N}}\right)={\textstyle \sum_{i=1}^{N}}J_{\underline{a}_{i}}J_{\underline{b}_{i}}+{\textstyle \sum_{i\neq j}}J_{\underline{a}_{i}}J_{\underline{b}_{j}}=\\
{\textstyle =\sum_{i=1}^{Q\left(\underline{a},\underline{b}\right)}}J_{\underline{a}_{i}}^{2}+{\textstyle \sum_{i>Q\left(\underline{a},\underline{b}\right)}}J_{\underline{a}_{i}}J_{\underline{b}_{i}}+{\textstyle \sum_{i\neq j}}J_{\underline{a}_{i}}J_{\underline{b}_{j}}\label{eq:ss-1}
\end{multline}
then, taking the expectation,
\begin{multline}
\mathbb{E}\left[M_{J}\left(\underline{a}\right)M_{J}\left(\underline{b}\right)\right]={\textstyle \sum_{i=1}^{Q\left(\underline{a},\underline{b}\right)}}\mathbb{E}\,(J_{\underline{a}_{i}}^{2})+\\
+{\textstyle \sum_{i>Q\left(\underline{a},\underline{b}\right)}\mathbb{E}}\,(J_{\underline{a}_{i}}J_{\underline{b}_{i}})+{\textstyle \sum_{i\neq j}}\mathbb{E}\,(J_{\underline{a}_{i}}J_{\underline{b}_{j}})={\textstyle \sum_{i=1}^{Q\left(\underline{a},\underline{b}\right)}}\gamma_{\underline{a}_{i}}^{2}\label{eq:ss}
\end{multline}
and if $\mathbb{E}\,(J_{\underline{a}_{i}}^{2})=\mathbb{E}\,(J_{a_{1}a_{2}...a_{i}}^{2})=\gamma_{\underline{a}_{i}}^{2}=\gamma_{\underline{0}_{i}}^{2}=\gamma_{i}^{2}$,
i.e., if we take 
\begin{equation}
J_{a_{1}a_{2}...a_{i}}=J_{i}
\end{equation}
in distribution with $J_{i}$ centered at zero with variance $\gamma_{i}^{2}$,
then 
\begin{equation}
\mathbb{E}\left[M_{J}\left(\underline{a}\right)M_{J}\left(\underline{b}\right)\right]={\textstyle \sum_{i=1}^{\,Q\left(\underline{a},\underline{b}\right)}}\gamma_{i}^{2}
\end{equation}
so that the matrix of the averaged products is ultrametric, see Figure 6.

\begin{figure}
\begin{centering}
\includegraphics[scale=0.14]{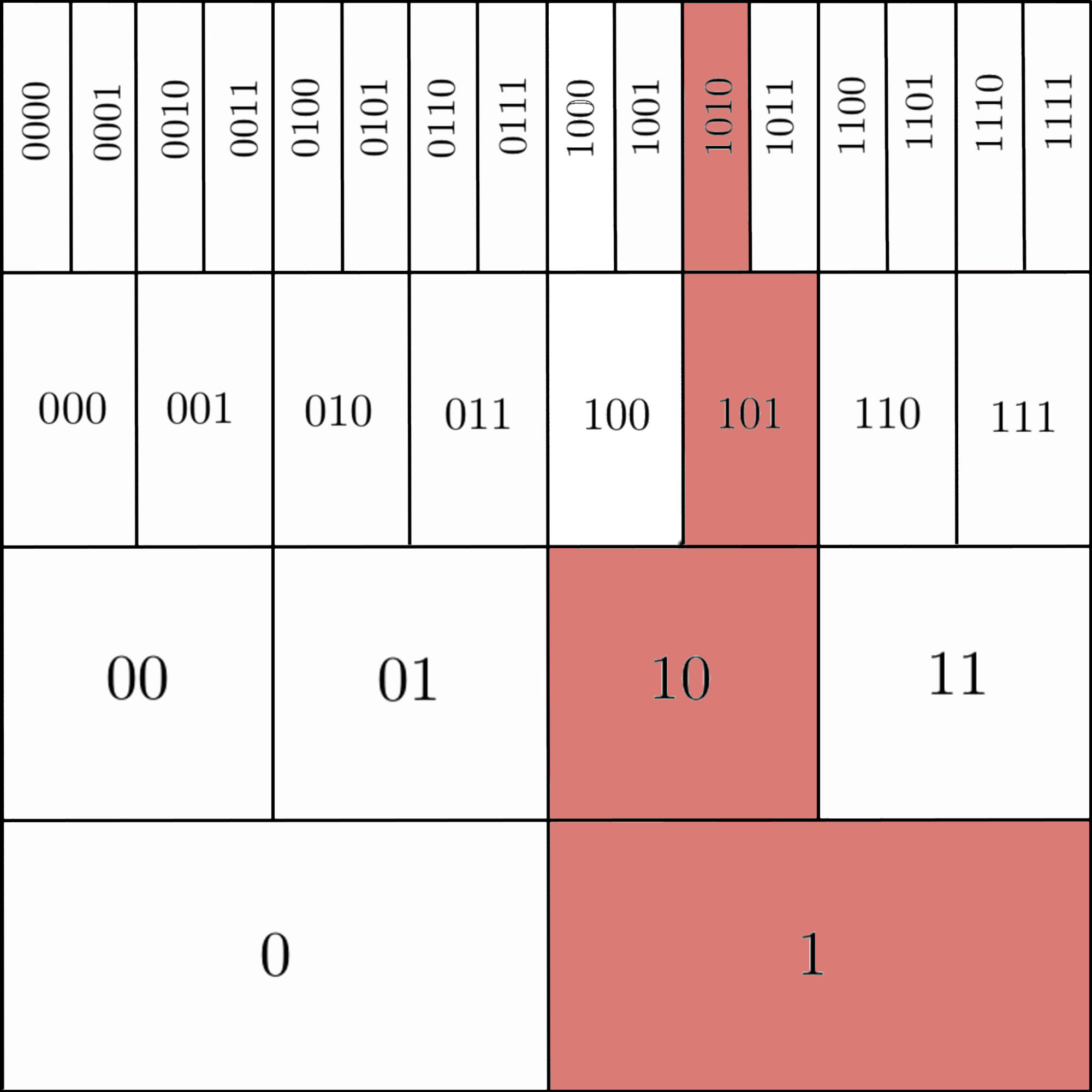}
\par\end{centering}
\centering{}\caption{Base kernel $A$. The kernel $A:\left[0,1\right]^{2}\rightarrow\mathbb{R}$
is subdivided into zones organized according to a tree-indexed partition;
$\left(1010\right)_{2}=\left(10\right)_{10}$ is shown. For each index
vector $\underline{a}_{i}=a_{1}a_{2}...a_{i}$ there is a corresponding
value $A\left(\underline{a}_{i}\right)$. If we take $A\left(\underline{a}_{i}\right)=J_{\underline{a}_{i}}=J_{a_{1}a_{2}...a_{i}}$
with $J_{\underline{a}_{i}}$ independent Gaussians of mean zero and
variance $\gamma_{i}^{2}$, the resulting kernel is ultrametric (the
scalar products of the columns are ultrametric on average).}
\end{figure}

\begin{figure}
\centering{}\includegraphics[scale=0.14]{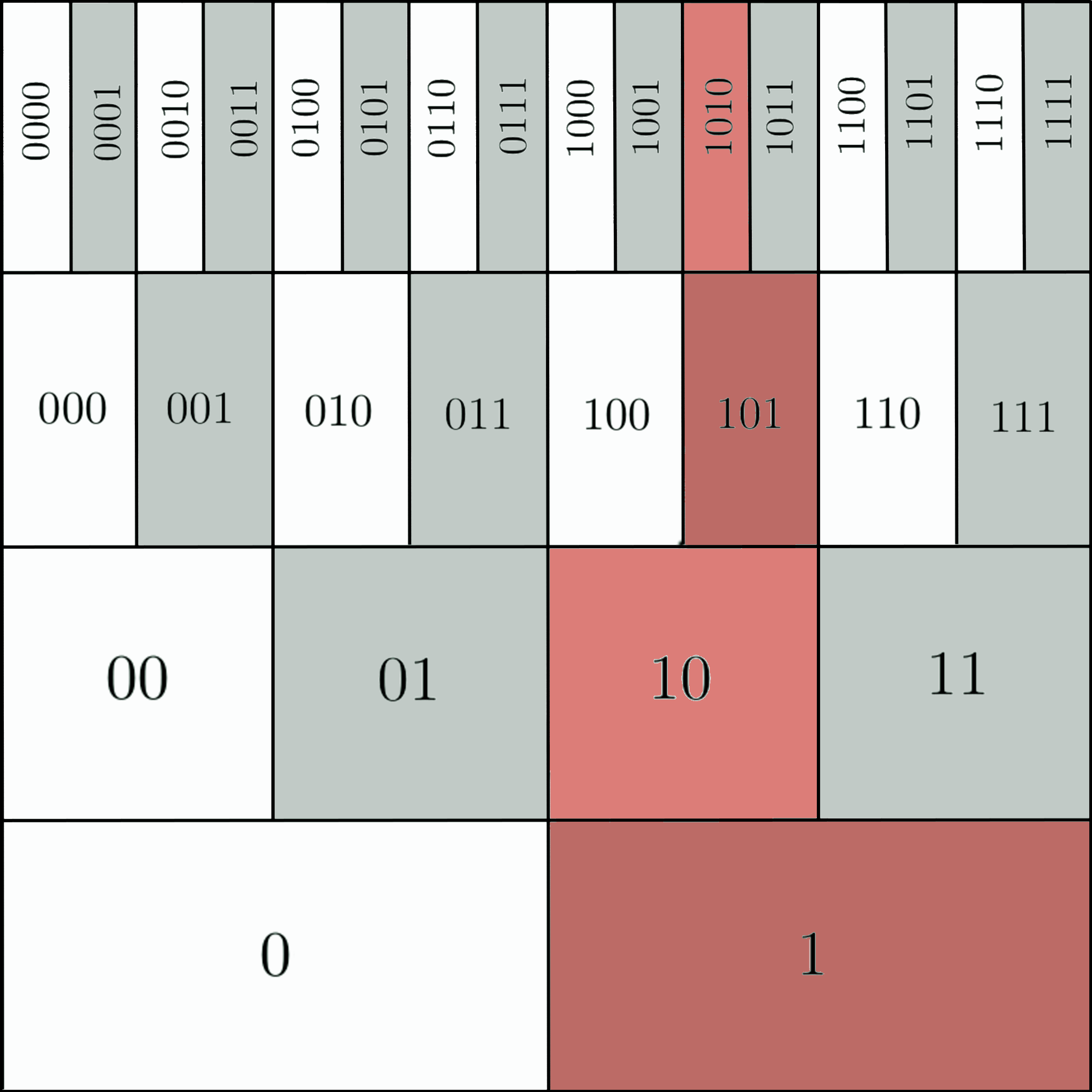}\caption{Choosing $A\left(\underline{a}_{i}\right)=2^{N-i}a_{i}$ gives the
binary numbering back. The pattern in the previous figure is the number
$\left(1010\right)_{2}=\left(10\right)_{10}$.}
\end{figure}

\begin{figure}
\begin{centering}
\includegraphics[scale=0.14]{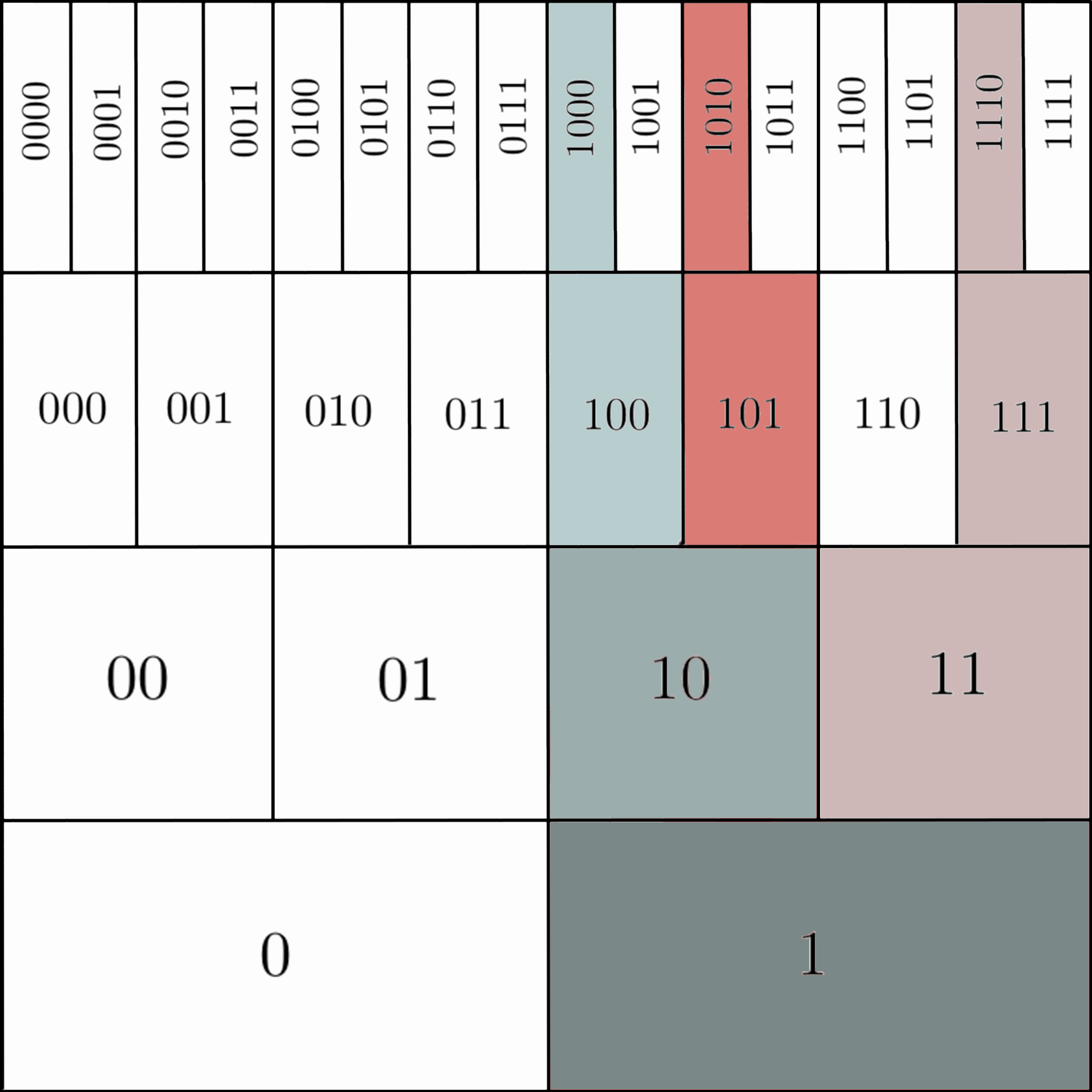}
\par\end{centering}
\centering{}\caption{Numbers in binary notation $\left(1010\right)_{2}=\left(10\right)_{10}$,
$\left(1000\right)_{2}=\left(8\right)_{10}$, $\left(1110\right)_{2}=\left(14\right)_{10}$.
The first two overlap in the first two digits. Then, both overlap
with the third in the first digit only. If we take $A\left(\underline{a}_{i}\right)=J_{a_{1}a_{2}...a_{i}}$
to be independent Gaussian random variables, the table of averaged products between two generic numbers
is exactly the expected, unnormalized scalar-product matrix of the kernel columns (overlap
matrix), which is also ultrametric if we take $J_{a_{1}a_{2}...a_{i}}=J_{i}$
in distribution.}
\end{figure}

\section{Canonical representation}

From there we can further generalize by introducing a variable base
numbering, where the base is an $N-$dimensional prime vector $\underline{p}$.
For a vector base
\begin{equation}
\underline{p}=\left\{ p_{i}\in\mathbb{P}:1\leq i\leq N\right\} ,
\end{equation}
the number of distinct numbers that we can write in such a base is $n\left(\underline{p}\right)$
\begin{equation}
n\left(\underline{p}\right)={\textstyle \prod_{i=1}^{N}}p_{i}
\end{equation}
Let the product of the prime sub-spaces
\begin{equation}
\mathbb{N}_{\underline{p}}={\textstyle \prod_{i=1}^{N}}\mathbb{N}_{p_{i}}
\end{equation}
be the space of $N-$digit numbers in $\underline{p}$ base 
\begin{equation}
\underline{a}=a_{1}\,a_{2}\,...\,a_{N}\in\mathbb{N}_{\underline{p}}
\end{equation}
where the $i-$th digit in the $\underline{p}$ base is
\begin{equation}
a_{i}\in\mathbb{N}_{p_{i}}=\left\{ 0,1,2\,...\,,p_{i}-1\right\} .
\end{equation}
 Let us define the partial products $s_{0}=1$ and 
\begin{equation}
s_{i}={\textstyle \prod_{j=1}^{i}}p_{N-j+1}.
\end{equation}
There is a map to the numbers $m$ from $0$ up to $n\left(\underline{p}\right)-1$
\begin{equation}
\mathbb{N}_{\underline{p}}\ni\underline{a}\ {\textstyle \overset{M_{\underline{p}}}{\longrightarrow}}\ m\in\mathbb{N}_{n\left(\underline{p}\right)}
\end{equation}
\begin{equation}
\mathbb{N}_{\underline{p}}\ni\underline{a}\ {\textstyle \overset{M_{\underline{p}}^{-1}}{\longleftarrow}}\ m\in\mathbb{N}_{n\left(\underline{p}\right)}
\end{equation}
The map $m\left(\underline{a}\right)=M_{\underline{p}}\left(\underline{a}\right)$
from $\underline{a}$ to $m$ is given by the formula \cite{KhrennikovKozyrev2006}
\begin{equation}
M_{\underline{p}}\left(\underline{a}\right)=s_{N-1}a_{1}+s_{N-2}a_{2}+\,...\,+s_{N-i}a_{i}+...\,+a_{N}
\end{equation}
from which we can write the first $n\left(\underline{p}\right)$ natural
numbers; the $p-$adic case is recovered for $p_{i}=p$. Notice that
each prime number $p_{i}$ can be itself mapped on a tree, a binary
one for example; this will be explored elsewhere.\footnote{We expect that adjusting the functional order parameter in the Parisi
theory of the SK model amounts to finding a suitable $\underline{p}-$base
for the problem; also, we expect that the branching ratio of the Ruelle
Cascade is the continuous analogue to that of a random $\underline{p}$
base; this should be formalized in some way. } 

\section{Fundamental theorem of arithmetic}

The fundamental theorem of Arithmetic guarantees that each positive
integer admits a decomposition into primes, i.e., a canonical representation,
that is unique up to commutations of factors in the product. Let $\mathbb{P}\subset\mathbb{N}$
be the set of prime numbers,
\begin{equation}
n={\textstyle \prod_{\,\,p\in\mathbb{P}}}\,p^{v_{n}\left(p\right)}
\end{equation}
where $v_{n}\left(p\right)$ is the multiplicity of the factor $p$
(that can be zero if $p$ is not a factor of $n$). We call
the vector 
\begin{equation}
v_{n}=\left\{ v_{n}\left(p\right)\in\mathbb{N}:\,p\in\mathbb{P}\right\} 
\end{equation}
the prime spectrum of $n$, which is unique for each $n$. We define the
prime support of $n$ 
\begin{equation}
\mathbb{P}_{n}=\left\{ p\in\mathbb{P}:\,v_{n}\left(p\right)>0\right\} 
\end{equation}
that is the set of prime factors of $n$. 

For each $n$ we can associate a $\underline{p}$--base, which we
indicate with $\underline{p}\left(n\right)$, with branching ratios
given by the prime factors of $n$ and multiplicities given
by the spectrum $v_{n}$. Consider the moments of the prime spectrum
of $n$ 
\begin{equation}
R_{\gamma}\left(n\right)={\textstyle \sum_{\,p\in\mathbb{P}}}\,v_{n}\left(p\right)^{\gamma}
\end{equation}
with $\gamma\in\left[0,1\right]$. For $\gamma=1$, this quantity
equals the number of prime factors of $n$, counted with multiplicity, and will be interpreted as the
total number of bosons 
\begin{equation}
R_{1}\left(n\right)={\textstyle \sum_{\,p\in\mathbb{P}}}\,v_{n}\left(p\right)
\end{equation}
while for $\gamma=0$ one gets the number of prime divisors of $n$
\begin{equation}
R_{0}\left(n\right)={\textstyle \sum_{\,p\in\mathbb{P}}}\,\mathbb{I}\left(v_{n}\left(p\right)\ge1\right)
\end{equation}
that can be identified with the number of types of bosons necessary
to get an excitation of energy $\log n$. 

\section{Example: Primon gas}

\begin{figure}[p]
\centering{}\includegraphics[scale=0.14]{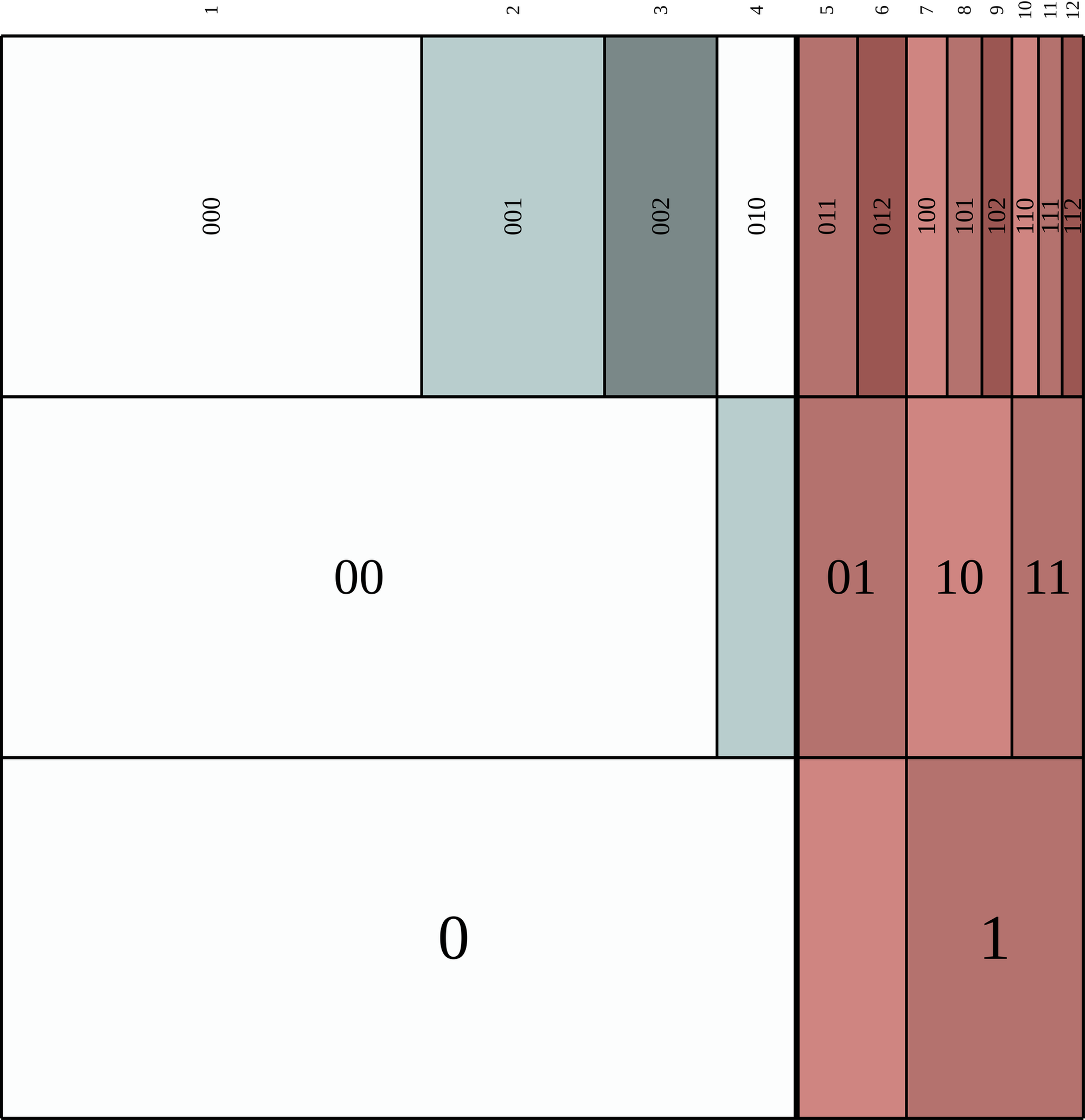}\caption{\protect\label{fig:Primon-kernel}Mixed-base approximation of the Primon Gibbs kernel for $L=2$ and $\beta=1.2$.
The bounds $v^{*}(2)=2$ and $v^{*}(3)=1$ determine the common base $(2,2,3)$,
with $N=3$ digits. From bottom to top, the first two rows contain binary digits
and the third contains ternary digits. Labels inside the blocks show the
successive prefixes, producing two, four and twelve blocks at the three levels.
White, light blue-grey and dark blue-grey encode the last digit values 0, 1 and 2,
respectively. The positive integer labels above the columns are obtained by
adding one to the mixed-base map of Eq. (31). Column widths follow the Gibbs
weights of Eq. (44), renormalized over the twelve displayed integers; the width
of each prefix block is the sum of the widths of its descendant columns.
The thick vertical line after 4 separates the regular set from the region
beginning at the first gap, $p_{3}=5$, highlighted by the translucent red overlay.
Below this threshold, the consecutive mixed-base numbering agrees with the
integers admitted by the bounded prime spectrum. Beyond it, the mixed-base
numbering continues consecutively, whereas the available prime factors and
their bounded multiplicities admit only 6 and 12. The red region therefore
identifies where the correspondence is no longer guaranteed. Equation (49)
controls the Gibbs mass of this region and its decay as the number of retained
primes increases at fixed temperature, or as the temperature decreases at fixed
truncation.}
\end{figure}

We conclude by recalling a physical interpretation of the Riemann
Zeta Function (RZF) due to Spector and Julia, \cite{Julia,Spector},
the so-called ``Primon'' gas. In this work, we extend this picture
by constructing a kernel representation of the Primon gas based on
a finite $p$-base, providing a concrete numerical framework for its
spectrum. Let $n\geq1$ be a positive integer and let $\mathbb{P\subset\mathbb{N}}$
be the set of prime numbers. Since prime numbers are numerable we can
keep the label $\ell$ to control the set $\mathbb{P}$; we denote by
$p_{\ell}$ the $\ell-$th prime number according to the chosen index.
We consider an index where the primes are ordered in increasing size,
$p_{\ell+1}>p_{\ell}$. By the fundamental theorem of arithmetic,
there is only one spectrum $v_{n}$ associated to $n$ such that
\begin{equation}
n={\textstyle \prod_{\,\ell\geq1}}\,p_{\ell}^{\,v_{n}\left(p_{\ell}\right)},
\end{equation}
Taking the logarithm of the previous formula one finds \cite{Julia,Spector,Weiss2004}
\begin{equation}
\log\left(n\right)={\textstyle \sum_{\,\ell\geq1}}\,v_{n}\left(p_{\ell}\right)\log p_{\ell}
\end{equation}
Writing $H\left(x\right)\equiv\log x$, we obtain 
\begin{equation}
H\left(n\right)={\textstyle \sum_{\,\ell\geq1}}\,v_{n}\left(p_{\ell}\right)H\left(p_{\ell}\right).
\end{equation}
For $\beta>1$, the canonical partition function of such a Hamiltonian
is the Riemann Zeta Function (RZF); this can be shown analytically
\begin{equation}
{\textstyle \sum_{\,n\geq1}}\exp\left[-\beta H\left(n\right)\right]={\textstyle \sum_{\,n\geq1}}n^{-\beta}=\zeta\left(\beta\right)
\end{equation}
but notice that the same result is obtained by physical arguments
if one assumes that the Hamiltonian represents a mixture of non-interacting
bosons 
\begin{multline}
\zeta\left(\beta\right)={\textstyle \sum_{\,n\geq1}}{\textstyle \prod_{\,\ell\geq1}}\exp\left[-\beta\,v_{n}\left(p_{\ell}\right)H\left(p_{\ell}\right)\right]=\\
={\textstyle \prod_{\,\ell\geq1}}{\textstyle {\textstyle \sum_{\,v\geq0}}}\exp\left[-\beta\,v\,H\left(p_{\ell}\right)\right]=\\
={\textstyle \prod_{\,\ell\geq1}}\left\{ 1-\exp\left[-\beta H\left(p_{\ell}\right)\right]\right\} ^{-1}.\label{eq:sd}
\end{multline}
The boson types are labeled by $\ell$ and represent the prime numbers;
each $n\in\mathbb{N}$ is then interpreted as an excitation of this
gas with spectrum $v_{n}\left(p_{\ell}\right)$, that is the prime
spectrum of $n$. In the second line of the above equation we sum
over all possible values of the prime spectrum $v_{n}\left(p_{\ell}\right)$,
which are interpreted as occupation numbers of the $\ell-$th boson
type. One can associate a free energy to such a gas
\begin{equation}
\beta f\left(\beta\right)={\textstyle \sum_{\,\ell\geq1}}\log\left\{ 1-\exp\left[-\beta H\left(p_{\ell}\right)\right]\right\} =-\log\zeta\left(\beta\right)
\end{equation}
and the relative Gibbs measure, that is 
\begin{equation}
\mu\left(\beta,n\right)=\frac{\exp\left[-\beta H\left(n\right)\right]}{\zeta\left(\beta\right)}=\frac{n^{-\beta}}{\zeta\left(\beta\right)}\label{eq:probab}
\end{equation}
which represents the probability of finding an excitation with spectrum
$v_{n}\left(p_{\ell}\right)$ in a Primon gas at temperature $1/\beta$.
This is interesting because we can quantify the probability mass
outside the first $n$ energy levels, and find that it
decays both in $n$ and in the inverse temperature $\beta$. By combining this
with the argument of the previous sections we can actually construct
a kernel representation of the Primon gas that accounts for a large
portion of the equilibrium ensemble. For $L\geq2$, consider a truncated spectrum
\begin{equation}
v_{n}:=\left\{ v_{n}\left(p_{1}\right),\,v_{n}\left(p_{2}\right),\,...\,,\,v_{n}\left(p_{L}\right)\right\} 
\end{equation}
containing only the first $L$ primes. By definition, with the first
$L$ primes we can write any number smaller than the $(L+1)-$th prime.
Let us introduce a notation for the maximum of each component with respect
to the condition $n<p_{L+1}$
\begin{equation}
v^{*}\left(p_{\ell}\right):={\textstyle \sup_{\ 1\leq n<p_{L+1}}}v_{n}\left(p_{\ell}\right)
\end{equation}
Then we could use $v^{*}$ to construct a common kernel base for any
integer less than $p_{L+1}$ without gaps. This is obtained by interpreting
the $\ell-$th component of the bound $v^{*}\left(p_{\ell}\right)$
as the multiplicity of $p\lyxmathsym{\textendash}$adic digits associated
to the prime $p_{\ell}$ in a $\underline{p}$--base with
\begin{equation}
N={\textstyle \sum_{\,\,\ell\leq L}}\,v^{*}\left(p_{\ell}\right)
\end{equation}
digits in total. We can bound the spectrum as follows
\begin{equation}
v^{*}\left(p_{\ell}\right)<\frac{\log p_{L+1}}{\log p_{\ell}}
\end{equation}
and still represent the first $p_{L+1}-1$ positive integers, together with zero, without
any gap, since the bound was chosen such that the missing integers
are greater than or equal to $p_{L+1}$. Hence, the mixed-base numbering agrees with the integers admitted by the
bounded prime spectrum throughout the regular set below the first gap.
Starting at the first gap, this correspondence is no longer guaranteed (Figure~\ref{fig:Primon-kernel}). This modified spectrum $v^{*}$ describes
an approximate Primon gas where some (in fact most) energy levels
labeled by integers larger than $p_{L+1}-1$ are neglected: this is closely related to the ``regular
set'' described in \cite{FranchiniRSBwR2023,Franchini2017,Franchini2016}.
By Eq. (\ref{eq:probab}), the Gibbs probability of the region where the
representations may differ satisfies
\begin{equation}
{\textstyle \sum_{\,n\geq p_{L+1}}}\,\mu\left(\beta,n\right)\leq\frac{\left(p_{L+1}-1\right)^{1-\beta}}{\left(\beta-1\right)\zeta\left(\beta\right)},\qquad\beta>1.
\end{equation}

\clearpage
\clearpage

\noindent This bound vanishes as the number of retained primes increases at fixed
temperature, or as the temperature decreases at fixed truncation, thereby
controlling the Gibbs mass associated with the discrepancy. Further investigations
will be presented elsewhere.

\section{Conclusions}

We have shown that p--adic theory, Replica Symmetry Breaking, and
kernel methods share a common formal structure. The generalization
to the mixed $p-$base allows any natural number to be represented
as a hierarchical spin state, while the construction of a $p-$base
for the Primon gas provides a concrete and controlled kernel realization
of the Riemann Zeta function. This result makes the connection between
number theory and kernel theory explicit and operational, offering
a novel framework to investigate arithmetic models through ultrametric
and kernel-based approaches. The kernel-based approach also provides a direct
connection to the statistical mechanics of networks, including Boltzmann machines
and the perceptron \cite{CCMBoltzmannMachines,CCMIsingPerceptron}, and to the
recently developed lattice field theory approach to natural neural networks
\cite{BardellaFranchini2024,BardellaFranchiniShort2024,FranchiniBardellaLATTICE2025}.

\section*{Acknowledgments}

The research presented in this work was conducted in the period 2016--2020
within the LoTGlasSy project (Parisi), funded by the European Research
Council (ERC) under the European Union\textquoteright s Horizon 2020
research and innovation program (Grant Agreement Num. 694925). We
are grateful to Giorgio Parisi (Accademia dei Lincei), Filippo Cesi
(Sapienza Università di Roma) and Nicola Kistler (Goethe University
Frankfurt) for interesting discussions.

\end{document}